\documentclass[aps,prd,superscriptaddress,twocolumn,floatfix,showpacs]{revtex4}

\usepackage{graphicx}
\def\prl{Phys. Rev. Lett.}
\def\prd{Phys. Rev. D}

\def\apj{Astrophys. J.}

\def\mnras{Mon. Not. R. Astron. Soc.}

\begin{document}
   
\title{A formalism for the construction of binary neutron stars with arbitrary circulation}

\author{Thomas W. Baumgarte}
\altaffiliation{Also at Department of Physics, University of Illinois at
  Urbana-Champaign, Urbana, IL 61801}
\affiliation{Department of Physics and Astronomy, Bowdoin College,
  Brunswick, ME 04011}
  
\author{Stuart L. Shapiro}
\altaffiliation{Also at Department of Astronomy and NCSA, University of
  Illinois at Urbana-Champaign, Urbana, IL 61801}
\affiliation{Department of Physics, University of Illinois at
  Urbana-Champaign, Urbana, IL 61801}

\begin{abstract}
Most numerical models of binary stars -- in particular neutron stars in compact binaries -- assume the companions to be either corotational or irrotational.  Either one of these assumptions leads to a significant simplification in the hydrodynamic equations of stationary equilibrium.   In this paper we  develop a new formalism for the construction of binary stars with circulation intermediate between corotational and irrotational.   Generalizing the equations for irrotational flow we cast the Euler equation, which is an algebraic equation in the case of corotational or irrotational fluid flow, as an elliptic equation for a new auxiliary quantity.  We also suggest a parameterized decomposition of the fluid flow that allows for a variation of the stellar circulation.
\end{abstract}

\maketitle

\section{Introduction}

Significant effort has gone into the numerical modeling of compact binaries containing neutron stars (see, e.g., \cite{BauS03} for a review).    Quasiequilibrium models of such binaries usually assume the existence of an approximate helical Killing vector that generates the binary orbit.  The existence of such an approximate Killing vector implies that, in a coordinate system corotating with the binary, the binary companions appear approximately stationary.   However, this notion leaves the rotation, or circulation, of the individual stars undetermined.  For example, each star could always show the same side to its companion, making it corotational, or it could have zero circulation as seen in an inertial frame, making it irrotational.  In principle, of course, each star could be rotating arbitrarily, about an arbitrary axis, and potentially even differentially.

As long as neutron stars can be modeled as ideal fluids, constructing neutron star solutions amounts to finding solutions to the Euler equation and the continuity equation.  For equilibrium solutions, the Euler equation simplifies in two special cases, namely for corotational and for irrotational binaries.  In both cases the Euler equation reduces to a total derivative and can be integrated once to yield an algebraic equation for the fluid enthalpy.  

For corotational binaries the continuity equation is satisfied identically, which simplifies the resulting equations significantly.   Not surprisingly, this assumption was adopted in the first quasiequilibrium studies of both binary neutron stars (see, e.g., \cite{BauCSST97,BauCSST98b}) and black hole-neutron star binaries (e.g.~\cite{BauSS04}).  Corotational flow does not seem very realistic, however, since tidally locking a neutron star in a compact binary would require an unphysically large viscosity (see \cite{Koc92,BilC92}).  

Instead, it seems more realistic to assume the fluid to be irrotational, which is a reasonable assumption as long as the orbital frequency is much larger than the spin frequency that the neutron stars would have in isolation.  In this case the velocity may be written as the gradient of a velocity potential.  The continuity equation then becomes an elliptic equation for this velocity potential.  This formalism was developed independently by \cite{BonGM97,Asa98,Teu98,Shi98} (see also \cite{Gou98} for a comparison between the different approaches), and numerical solutions have been constructed both for binary neutron stars (e.g.~\cite{BonGM99a,GouGTMB01,TanG02}, see also \cite{UryE99} for Newtonian models) and black hole-neutron star binaries (e.g.~\cite{TanBFS05,TanBFS08,Gra06}).

Constructing binaries containing rapidly rotating stars, e.g.~millisecond pulsars, requires a formalism that allows for arbitrary spin configurations.  In this case the Euler equation no longer yields a trivial first integral.   So far, neutron stars with arbitrary circulation have been constructed only by \cite{MarS03}, hereafter MS.  MS make several assumptions in their treatment of neutron stars with arbitrary circulation.  Instead of using the Euler equation, they solve the Bernoulli equation, which states that a certain quantity is constant along fluid trajectories.  MS then find approximate solutions by assuming that this quantity is constant globally, which turns out to be correct for corotational and irrotational fluid flow, but not in general.   While this approximation may be quite accurate, it is of interest, at the very least as a point of principle, to identify a formalism that avoids this approximation.  MS further assume that a solenoidal part of the fluid flow is centered on the stellar centers, determined by the point of maximum baryonic density.  This may also be a good approximation, and yields corotational and irrotational models in good agreement with other approaches, but again it would be desirable to develop a formalism that avoids this assumption.

In this paper we develop a formalism that improves on that of MS in several ways.  We suggest a way in which the Euler equation can be solved for circulation intermediate between corotational and irrotational, and specify the fluid flow in a geometrically motivated fashion that does not make any assumptions about the centers of rotation.  After stating some preliminaries in Section \ref{sec:prelims} we will, for the sake of clarity, first develop this formalism in a Newtonian framework in Section \ref{sec:Newt}.   We will then treat the problem in a relativistic framework in Section \ref{sec:Rel}, and will briefly summarize in Section \ref{sec:Sum}.  In Appendix \ref{AppA} we will briefly discuss an alternative decomposition of the fluid flow.

\section{Preliminaries}
\label{sec:prelims}

We adopt a 3+1 decomposition of the spacetime metric and write the spacetime metric $g_{ab}$ in the form
\begin{equation}
g_{ab} dx^a dx^b = - \alpha^2 dt^2 + \gamma_{ij} (dx^i + \beta^i dt) (dx^j + \beta^j dt).
\end{equation}
Here $\alpha$ is the lapse function, $\beta^i$ is the shift vector and $\gamma_{ij}$ is the spatial metric induced on a spatial slice $\Sigma$ of constant coordinate time $t$.  The normal on this slice is 
\begin{equation}
n^a = \alpha^{-1} ( 1, - \beta^i).
\end{equation}
Here and in the following we denote spacetime indices with letters $a,b,c,\ldots$ and spatial indices with letters $i,j,k,\ldots$.  The coordinate congruence $t^a$, which connects points with the same spatial coordinate labels on different slices $\Sigma$, is then given by $t^a = \alpha n^a + \beta^a$.

We also assume the existence of a helical Killing vector $\xi^a_{\rm hel}$, and assume it to be normalized in such a way that its time component is unity, $\xi^t_{\rm hel} = 1$.  We can write this Killing vector as
\begin{equation} \label{hel_def}
\xi^a_{\rm hel} = t^a + k^a,
\end{equation}
where the vector $k^a$ is purely spatial and describes the orbital rotation
\begin{equation}
k^a = \Omega \left( \frac{\partial}{\partial \phi} \right)^a.
\end{equation}
In cartesian coordinates, for a rotation about the $z$-axis, we may also write the spatial components of $k^a$ as
\begin{equation} \label{Newt:k_cart}
k^i = \Omega (-y,x,0).
\end{equation}

\section{Newtonian treatment}
\label{sec:Newt}

\subsection{Newtonian equations of stationary equilibrium}
\label{subsec:Newt:eq}

The Newtonian equations of hydrodynamics are the Euler equation
\begin{equation} \label{Newt:Euler0}
\partial_t v_i + v^j D_j v_i = - \rho_0^{-1} D_i P - D_i \phi_{\rm N}
\end{equation}
and the continuity equation
\begin{equation} \label{Newt:cont0}
\partial_t \rho_0 + D_i ( \rho_0 v^i) = 0.
\end{equation}
Here $v^i$ is the fluid velocity, $\rho_0$ the rest-mass density, $P$ the pressure, and $\phi_{\rm N}$ the Newtonian potential, which has to satisfy the Poisson equation
\begin{equation} \label{Newt:Poisson}
D^2 \phi_{\rm N} = 4 \pi \rho_0.
\end{equation}
We denote a covariant derivative with $D_i$, and the Laplace operator with $D^2 \equiv D^i D_i$.  It is also useful to define $V^i$ as the fluid velocity relative to $k^a$, 
\begin{equation} \label{Newt:V}
v^i = k^i + V^i,
\end{equation}
i.e.~$V^i$ is the velocity as seen in a frame corotating with the binary.  For isentropic flow we also have 
\begin{equation}
\rho_0^{-1} D_i P = D_i h.
\end{equation}
Here $h$ is the specific enthalpy, which, adopting a convention more common in a relativistic framework, is defined as 
\begin{equation}
h = 1 + \epsilon + P/\rho_0,
\end{equation}
where $\epsilon$ is the specific internal energy density.  We also assume that we are given an equation of state that relates the enthalpy $h$ to the density $\rho_0$.

For stationary solutions the Lie derivative of the hydrodynamic variables $\rho_0$ and $v^i$ has to vanish, i.e.
\begin{equation} \label{Newt:Lie_v}
{\mathcal L}_{\xi_{\rm hel}} v_i = \partial_t v_i + {\mathcal L}_{\bf k} v_i 
= \partial_t v_i + k^j D_j v_i - v_j D_i k^j = 0
\end{equation}
and similar for $\rho_0$.  Combining equation (\ref{Newt:Lie_v}) with the Euler equation (\ref{Newt:Euler0}) we then obtain
\begin{equation} \label{Newt:Euler}
D_i \left( h - \frac{1}{2} v^2 + v_j V^j + \phi_{\rm N} \right)
+ V^j ( D_j v_i - D_i v_j ) = 0,
\end{equation}
while the equation of continuity becomes
\begin{equation} \label{Newt:cont}
D_i (\rho_0 V^i) = 0.
\end{equation}
Equations (\ref{Newt:Euler}) and (\ref{Newt:cont}) form the Newtonian equations of stationary equilibrium.

\subsection{Corotational and irrotational flow}
\label{subsec:Newt:coir}

It is easy to see how the above equations simplify for corotational or irrotational fluid flow.  

For corotational flow, we must have $V^i = 0$, in which case the continuity equation (\ref{Newt:cont}) is satisfied identically.  Furthermore, the second term in the Euler equation (\ref{Newt:Euler}) vanishes identically, so that the first term can be integrated to yield an algebraic equation for the enthalpy $h$,
\begin{equation} \label{Newt:Euler_corot}
h - \frac{1}{2} v^2 + \phi_{\rm N} = C.
\end{equation}
Here $C$ is some constant of integration, and $v^2$ can be found from (\ref{Newt:V}), which now implies $v^i = k^i$.  Here and below we assume that both stars are identical for simplicity; the method is easily generalized for non-identical stars.

For irrotational flow, we require the fluid's vorticity to vanish, which is automatically the case if we write the fluid velocity as a gradient of a velocity potential $\Phi$,
\begin{equation} \label{Newt:v_irrot}
v_i = D_i \Phi.
\end{equation}
In this case, the second term in the Euler equation (\ref{Newt:Euler}) also vanishes identically, so that we again obtain an algebraic equation for the enthalpy $h$,
\begin{equation} \label{Newt:Euler_irrot}
h - \frac{1}{2} v^2 + v_j V^j + \phi_{\rm N} = C.
\end{equation}
The continuity equation (\ref{Newt:cont}) now becomes an elliptic equation for the velocity potential $\Phi$,
\begin{equation} \label{Newt:Phi_irrot_eq}
D^2 \Phi = (k^i - D^i \Phi) D_i \ln \rho_0.
\end{equation}
To ensure that the right hand side remains finite at the stellar surface, where $\rho_0$ becomes zero, we must have
\begin{equation} \label{Newt:Phi_irrot_bound}
\left. (k^i - D^i \Phi) D_i \rho_0 \right|_{\rm surf} = 0.
\end{equation}
Given that $D_i \rho_0$ is normal to the stellar surface, this relation furnishes a Neumann boundary condition for $\Phi$.   We could have derived the same condition by requiring that, at the stellar surface, the fluid flow be tangent to the surface.

\subsection{Intermediate circulation}

For intermediate circulation, the second term in the Euler equation (\ref{Newt:Euler}) no longer vanishes, and we can no longer find a simple first integral.   One approach to find a solution is to take a divergence of the equation, which yields an elliptic equation 
\begin{equation} \label{Newt:H1}
D^2 H + D^i \left( V^j (D_j v_i - D_i v_j) \right) = 0
\end{equation}
for a new auxiliary quantity 
\begin{equation} \label{Newt:Hdef}
H \equiv h - \frac{1}{2} v^2 + v_j V^j + \phi_{\rm N}.
\end{equation}
In the following we will interpret this as an equation for the specific enthalpy $h$,
\begin{equation}\label{Newt:h_eq}
h = H + \frac{1}{2} v^2 - v_j V^j - \phi_{\rm N}.
\end{equation}

Before proceeding we need to specify the circulation of the fluid.  To do so, we choose to decompose the 
velocity $v_i$ according to
\begin{equation} \label{Newt:v}
v_i = D_i \Phi + \eta k_i,
\end{equation}
which, from (\ref{Newt:V}), implies
\begin{equation} \label{Newt:V_ans}
V^i = D^i \Phi + (\eta - 1) k^i.
\end{equation}
Here $\Phi$ is a generalized velocity potential, and $\eta$ is an arbitrary constant.  As we will discuss in Section \ref{subsec:Newt:limits} we recover irrotational flow for $\eta = 0$ and corotational flow for $\eta = 1$.  With our decomposition (\ref{Newt:v}), $\eta$ therefore parameterizes a sequence of stellar models connecting corotational and irrotational models.   By generalizing the decomposition  (\ref{Newt:v}) it should be possible to change the axis of the internal rotation, or change the degree of differential rotation.
 
Inserting the decomposition (\ref{Newt:v}) into equation (\ref{Newt:H1}) we first note that the velocity potential $\Phi$ again drops out of the term
\begin{equation}
D_j v_i - D_i v_j = \eta \,(D_j k_i - D_i k_j) \equiv \eta \, \kappa_{ij},
\end{equation}
where we have defined $\kappa_{ij}$ for convenience.  Evidently, $\kappa_{ij}$ is antisymmetric by construction and can be computed directly from (\ref{Newt:k_cart}).   The second term in equation (\ref{Newt:H1}) then becomes
\begin{equation}
D^i \left( V^j (D_j v_i - D_i v_j) \right) = 
\eta \left( (D^i V^j) \kappa_{ij} + V^j D^i \kappa_{ij} \right). 
\end{equation}
Inserting (\ref{Newt:V_ans}) to evaluate the first term on the right hand side we find
\begin{eqnarray}
(D^i V^j)  \kappa_{ij} & = & (D^i D^j \Phi + (\eta - 1) D^i k^j) \kappa_{ij} \nonumber \\ 
& = & (\eta - 1) (D^i k^j) \kappa_{ij},
\end{eqnarray}
where the second equality holds because $D^i D^j \Phi$ is symmetric while $\kappa_{ij}$ is antisymmetric.  That means that all second derivatives of $\Phi$ disappear in equation (\ref{Newt:H1}), leaving us with
\begin{equation} \label{Newt:H_eq}
D^2 H + \eta (\eta - 1) (D^i k^j) \kappa_{ij} + \eta \left( D^j \Phi + (\eta - 1) k^j \right) D^i \kappa_{ij} = 0.
\end{equation}
This is the form of the Euler equation that generalizes equations (\ref{Newt:Euler_corot}) or (\ref{Newt:Euler_irrot}) for intermediate circulation.  Instead of an algebraic equation for the enthalpy $h$ we now have to solve an elliptic equation for the new auxiliary quantity $H$.  Given solutions for the other fields we can then find the enthalpy $h$ from (\ref{Newt:h_eq}).  On the stellar surface the density vanishes, meaning that $h$ approaches unity.  This requirement leads to the Dirichlet boundary condition
\begin{equation} \label{Newt:H_bound}
\left. H \right|_{\rm surf} = 1 - \frac{1}{2} v^2 + v_j V^j + \phi_{\rm N}
\end{equation}
on the stellar surface.

To find an equation for the velocity potential $\Phi$ we insert (\ref{Newt:V_ans}) into the continuity equation (\ref{Newt:cont}), which yields
\begin{equation} \label{Newt:Phi_eq}
D^2 \Phi = - \left( D^i \Phi + (\eta - 1) k^i \right) D_i \ln \rho_0.
\end{equation}
To ensure that the right hand side remains finite on the stellar surface we must have
\begin{equation} \label{Newt:Phi_bound}
\left. \left( D^i \Phi + (\eta - 1) k^i \right) D_i \rho_0 \right|_{\rm surf} = 0.
\end{equation}
As in the case of irrotational flow, this conditions forms a Neumann boundary condition.

We have now assembled all equations and boundary conditions needed to construct binary stars with circulation intermediate between corotational and irrotational flow in a Newtonian framework.  In addition to the Poisson equation (\ref{Newt:Poisson}), we need to solve equation (\ref{Newt:H_eq}), subject to the boundary condition (\ref{Newt:H_bound}), for $H$, and equation (\ref{Newt:Phi_eq}), subject to boundary condition (\ref{Newt:Phi_bound}), for $\Phi$.  Given these solutions we can find $v_i$ and $V^i$ from equations (\ref{Newt:v}) and (\ref{Newt:V_ans}), and $h$ from (\ref{Newt:h_eq}).  Finally, $\rho_0$ can be computed from $h$ given an equation of state, and $\Omega$ is determined by enforcing a circular orbit (see, e.g., \cite{UryE99}). 

\subsection{Limiting cases}
\label{subsec:Newt:limits}

Before proceeding to a relativistic treatment of the problem it is useful to see how the equations for corotational and irrotational flow appear as limiting cases of the equations for intermediate circulation.  Clearly, our equations for intermediate circulation are consistent with equations (\ref{Newt:Euler}) and (\ref{Newt:cont}), and are therefore satisfied both for corotational flow with $V^i = 0$ and irrotational flow with $v_i = D_i \Phi$.  In the following we demonstrate that these solutions are also consistent with the decomposition (\ref{Newt:v}).  

We start with irrotational flow, which we obtain for $\eta = 0$.   In this limit, the decomposition (\ref{Newt:v}) evidently reduces to (\ref{Newt:v_irrot}), and both the equation (\ref{Newt:Phi_eq}) and boundary condition (\ref{Newt:Phi_bound}) reduce to their equivalents for irrotational flow, equations (\ref{Newt:Phi_irrot_eq}) and (\ref{Newt:Phi_irrot_bound}).  For $\eta = 0$, equation (\ref{Newt:H_eq}) reduces to a Laplace equation for $H$, which is uniquely solved by $H = C_{H}$, where $C_{H}$ is a constant, when $H = C_{H}$ on the boundary.  This is consistent with the other equations, which can be seen by evaluating equation (\ref{Newt:Euler_irrot}) on the stellar surface.  As a result, the equation for the enthalpy $h$ reduces to equation (\ref{Newt:Euler_irrot}), as expected.

We obtain corotational flow by setting $\eta = 1$.  To see this, we first observe that both the equation (\ref{Newt:Phi_eq}) for the velocity potential $\Phi$ and its boundary condition (\ref{Newt:Phi_bound}) are solved by $\Phi = C_{\Phi}$, where $C_{\Phi}$ is a constant.  Equation (\ref{Newt:V_ans}) then implies $V^i = 0$, as expected for corotational flow.  Equation (\ref{Newt:H_eq}) again reduces to a Laplace equation which, as in the irrotational case, is solved by $H = C_{H}$.  The equation for the enthalpy $h$ therefore reduces to equation (\ref{Newt:Euler_corot}).

\section{Relativistic treatment}
\label{sec:Rel}

\subsection{Relativistic equations of stationary equilibrium}
\label{subsec:rel:eq}

The relativistic equations of hydrodynamics follow from the conservation of energy-momentum,
\begin{equation} \label{eq:divT}
\nabla_b T^{ab} = 0,
\end{equation}
whose spatial components form the relativistic Euler equation, and the conservation of rest-mass
\begin{equation} \label{eq:divn}
\nabla_a ( \rho_0 u^a ) = 0.
\end{equation}
Here $T^{ab}$ is the fluid's energy-momentum tensor, $u^a$ its four-velocity, and $\nabla_a$ denotes a covariant derivative associated with the four-metric $g_{ab}$.  Following the approach of \cite{Shi98} we can write $u^a$ as
\begin{equation} \label{rel:V_def}
u^a = u^t (\xi^a_{\rm hel} + V^a),
\end{equation}
where the helical Killing vector $\xi^a_{\rm hel}$ was introduced in equation (\ref{hel_def}) and where $V^a$ is purely spatial, $n_a V^a = 0$.  Just like its Newtonian counterpart in equation (\ref{Newt:V}), $V^a$ now measures the fluid velocity in a frame comoving with the binary.  It is also convenient to define 
\begin{equation} \label{rel:u_def}
\hat u_i = \gamma_i{}^a h u_a.
\end{equation}
In terms of $\hat u^i$ the normalization of the four-velocity $u_a u^a = -1$  may be written as
\begin{equation} \label{rel:u_norm}
\alpha u^t = \left( 1 + h^{-2} \, \hat u_i \hat u^i \right)^{1/2}.
\end{equation}

For stationary solutions the Lie derivative of the hydrodynamical variables along $\xi^a_{\rm hel}$ again has to vanish.  As shown in \cite{Shi98}, the relativistic Euler equation then becomes
\begin{equation} \label{rel:Euler}
D_i \left( \frac{h}{u^t} + \hat u_j V^j \right) + V^j  \left( D_j \hat u_i - D_i \hat u_j \right) = 0,
\end{equation}
while the continuity equation becomes
\begin{equation} \label{rel:cont}
D_i \left( \alpha u^t \rho_0 V^i \right) = 0.
\end{equation}
Here $D_i$ is the covariant derivative associated with the spatial metric $\gamma_{ij}$.

We also assume that, on the slice $\Sigma$, the spatial metric $\gamma_{ij}$,  the lapse $\alpha$, and the shift $\beta^i$ are constructed by solving some subset of Einstein's equations, for example by using the conformal thin-sandwich decomposition.  These equations then play the role of the Poisson equation (\ref{Newt:Poisson}) in the Newtonian framework.  As before we further assume that we are given an equation of state that relates the specific enthalpy $h$ to the rest density $\rho_0$.

\subsection{Corotational and irrotational flow}
\label{subsec:rel:coir}

As in the Newtonian case it is easy to see how the equations simplify for both corotational and irrotational flow.

For corotational flow the four-velocity $u^a$ is aligned with the helical Killing vector $\xi^a_{\rm hel}$, so that $V^a = 0$.  The only remaining term in the Euler equation (\ref{rel:Euler}) can then be integrated to yield
\begin{equation} \label{rel:Euler_corot}
\frac{h}{u^t} = C,
\end{equation}
where $C$ is some constant, while the continuity equation (\ref{rel:cont}) is satisfied identically.

To obtain irrotational flow, with vanishing relativistic fluid vorticity, we now assume
\begin{equation} \label{rel:u_irrot}
\hat u_i = D_i \Phi,
\end{equation}
where $\Phi$ is a velocity potential (see \cite{Shi98} for a discussion).   With this choice the second term in the Euler equation (\ref{rel:Euler}) vanishes again, and we can find the first integral
\begin{equation} \label{rel:Euler_irrot}
\frac{h}{u^t} + \hat u_i V^i = C.
\end{equation}
Combining equations (\ref{rel:V_def}), (\ref{rel:u_norm}) and (\ref{rel:Euler_irrot}) we also find
\begin{equation} \label{rel:ut_irrot}
\alpha u^t = \frac{1}{\alpha h} (C + B^i D_i \Phi),
\end{equation}
where we have defined a rotational shift vector
\begin{equation} \label{rel:B_def}
B^a \equiv \beta^a + k^a.
\end{equation}
Using (\ref{hel_def}) and $t^a = \alpha n^a + \beta^a$ we may also write $B^a$ as
\begin{equation}
B^a = \xi^a - \alpha n^a.
\end{equation}
We will derive a more general version of equation (\ref{rel:ut_irrot}) for intermediate circulation below.

Note that $V^i$ can always be expressed as
\begin{equation} \label{rel:V_u}
V^i = \gamma^i{}_a V^a = 
\gamma^i{}_a \left( \frac{1}{u^t}  u^a - \xi^a_{\rm hel} \right) 
= \frac{1}{u^t h} \hat u^i - B^i.
\end{equation}
For irrotational flow we can now insert (\ref{rel:u_irrot}),
\begin{equation} 
V^i = \frac{1}{u^t h} D^i \Phi - B^i, \label{rel:V_irrot}
\end{equation}
and insert this result into the continuity equation (\ref{rel:cont}) to obtain
\begin{equation}
D_i (\alpha \rho_0 h^{-1} D^i \Phi) - D_i (\alpha u^t \rho_0 B^i) = 0.
\end{equation}
We may eliminate $u^t$ from this equation by inserting (\ref{rel:ut_irrot}), which yields
\begin{eqnarray} \label{rel:Phi_irrot_eq}
&& \!\!\!\! D^2 \Phi - D_i \left( \frac{C + B^j D_j \Phi}{\alpha^2} B^i \right) \nonumber \\
&& ~~~~ = \left( \frac{C + B^j D_j \Phi}{\alpha^2} B^i  - D^i \Phi \right) D_i \ln \frac{\alpha \rho_0}{h}.
\end{eqnarray} 
As expected, this equation reduces to (\ref{Newt:Phi_irrot_eq}) in the Newtonian limit.  A Neumann boundary condition again results from requiring that the right hand side remain finite at the stellar surface,
\begin{equation} \label{rel:Phi_irrot_bound}
\left. \left( \frac{C + B^j D_j \Phi}{\alpha^2} B^i  - D^i \Phi \right) D_i \frac{\alpha \rho_0}{h} 
\right|_{\rm surf} = 0.
\end{equation}
To complete the argument, we could insert equations (\ref{rel:u_irrot}), (\ref{rel:ut_irrot}) and (\ref{rel:V_irrot}) into (\ref{rel:Euler_irrot}) to obtain an equation for $h$ in terms of $\Phi$, and independently of $u^t$.

\subsection{Intermediate circulation}
\label{subsec:rel:arb_spin}

As in the Newtonian case, we observe that for intermediate circulation the second term in the Euler equation (\ref{rel:Euler}) no longer vanishes, so that we can no longer find a first integral.   Taking a divergence of the equation, we can again obtain an elliptic equation 
\begin{equation} \label{rel:H1}
D^2 H + D^i \left( V^j (D_j \hat u_i - D_i \hat u_j) \right) = 0
\end{equation}
for a new auxiliary function
\begin{equation} \label{rel:H_def}
H \equiv \frac{h}{u^t} + \hat u_i V^i.
\end{equation}
We will again interpret this as an algebraic equation for the specific enthalpy $h$.

Generalizing the Newtonian expression (\ref{Newt:v}) we decompose $\hat u_i$ according to
\begin{equation} \label{rel:u_ans}
\hat u_i = D_i \Phi + \eta k_i.
\end{equation}
Repeating the steps in equation (\ref{rel:V_irrot}) we now find
\begin{equation} \label{rel:V_ans}
V^i = \frac{1}{u^t h} (D^i \Phi + \eta k^i) - B^i.
\end{equation}
We point out that there are many different ways of generalizing the Newtonian decomposition (\ref{Newt:v}), and we will discuss an attractive alternative to the decomposition (\ref{rel:u_ans}) in Appendix \ref{AppA}.  Moreover, the Newtonian decomposition (\ref{Newt:v}) represents only one particular sequence that connects corotational and irrotational fluid flow; it should also be possible to generate models that rotate about an arbitrary axis of rotation.  Similarly, the relativistic decomposition (\ref{rel:u_ans}) represents only one possible sequence.  

Before proceeding we derive an equation for $u^t$.  Starting with the definition (\ref{rel:H_def}) we insert   equation (\ref{rel:V_u}) to obtain
\begin{equation}
\frac{h}{u^t} + \hat u_i \left(\frac{1}{u^t h} \hat u^i - B^i \right) = H,
\end{equation}
or
\begin{equation}
\frac{h}{u^t} \left(1 + h^{-2} \hat u_i \hat u^i \right) = H + \hat u_i B^i.
\end{equation}
We now insert the normalization (\ref{rel:u_norm}) as well as the decomposition (\ref{rel:u_ans}) to find
\begin{equation} \label{rel:ut}
\alpha u^t = \frac{1}{\alpha h} \left(H + B^i (\eta k_i + D_i \Phi) \right).
\end{equation}
For the irrotational case, when $\eta = 0$ and $H = C$, we recover equation (\ref{rel:ut_irrot}), as expected. 

Returning to the Euler equation (\ref{rel:H1}), we observe that
\begin{equation}
D_j \hat u_i - D_i \hat u_j = \eta \, (D_j k_i - D_i k_j) \equiv \eta \, \kappa_{ij},
\end{equation}
in complete analogy to the Newtonian case.  Expanding the terms in (\ref{rel:H1}) we now find
\begin{eqnarray} \label{rel:H_eq}
D^2 H \! \! & + \! \! &  \eta \left( (D^j \Phi + \eta k^j) D^i \frac{1}{u^t h} + \frac{\eta}{u^t h} D^i k^j - D^i B^j \right) \kappa_{ij}  \nonumber \\
\! \! & + \! \!  & \eta \left( \frac{1}{u^t h} (D^j \Phi + \eta k^j ) - B^j \right) D^i \kappa_{ij} = 0.
\end{eqnarray}
If desired, equation (\ref{rel:ut}) can be inserted to eliminate the terms $u^t h$.   Requiring that the density vanish on the stellar surface, so that $h = 1$, again leads to a Dirichlet boundary condition
\begin{equation} \label{rel:H_bound}
\left. H \right|_{\rm surf} =  \frac{1}{u^t} + \hat u_i V^i.
\end{equation}

We now turn to the continuity equation (\ref{rel:cont}).  Inserting the decomposition (\ref{rel:V_ans}), we obtain
\begin{equation}
D_i \left(\frac{\alpha \rho_0}{h} (D^i \Phi + \eta k^i) \right) - D_i (\alpha u^t \rho_0 B^i) = 0,
\end{equation}
or, eliminating $u^t$ with the help of equation (\ref{rel:ut}),
\begin{eqnarray} \label{rel:Phi_eq}
&& \! \! D^2 \Phi - D_i \left( \frac{H + B^j (\eta k_j + D_j \Phi)}{\alpha^2} B^i \right) = - \eta D_i k^i  \\
& & \! \!  ~~ + \left(\frac{H + B^j (\eta k_j + D_j \Phi)}{\alpha^2} B^i - D^i \Phi - \eta k^i \right) 
D_i \ln \frac{\alpha \rho_0}{h}  \nonumber  
\end{eqnarray}
As before we require that the right hand side remain finite at the stellar surface, which yields the boundary condition
\begin{equation} \label{rel:Phi_bound}
\left. \left(\frac{H + B^j (\eta k_j + D_j \Phi)}{\alpha^2} B^i - D^i \Phi - \eta k^i \right) 
D_i \frac{\alpha \rho_0}{h} \right|_{\rm surf} = 0.
\end{equation}

We have now assembled a complete set of equations and boundary conditions that can be used to construct relativistic binary stars with intermediate circulation.  Given a choice for the parameter $\eta$ we can solve equation (\ref{rel:H_eq}) for $H$ and equation (\ref{rel:Phi_eq}) for $\Phi$.  We can then find
the combination $u^t h$ from (\ref{rel:ut}), and the velocity expressions $V^i$ and $\hat u_i$ from equations (\ref{rel:u_ans}) and (\ref{rel:V_ans}).  Finally, $h$ can be found from the definition (\ref{rel:H_def}), $\rho_0$ from the equation of state, and $\Omega$ can be determined by enforcing a circular orbit (see \cite{BauS03} and references therein).   The above equations have to be solved together with gravitational field equations of the lapse $\alpha$, the shift $\beta^i$ (which, together with the helical Killing vector (\ref{hel_def}), determines the rotational shift vector $B^i$) and the spatial metric $\gamma_{ij}$; this can be accomplished, for example, by solving the constraint equations of Einstein's field equations in the conformal thin-sandwich decomposition.  All the above equations are coupled in complicated ways, and presumably the easiest way of constructing numerical solutions involves an iterative algorithm.

\subsection{Limiting cases}
\label{subsec:rel:limits}

Our equations are based on the Euler equation (\ref{rel:Euler}) and the continuity equation (\ref{rel:cont}), which are clearly satisfied for both corotational flow with $V^i = 0$ and irrotational flow with $\hat u_i = D_i \Phi$.  However, it is less clear whether these solutions are consistent with our decomposition (\ref{rel:u_ans}) and (\ref{rel:V_ans}).

As in the Newtonian case we can recover irrotational flow for $\eta = 0$.  In this limit, all equations, including a constant solution $H = C_H$ for $H$, are consistent with the corresponding equations for irrotational flow in Section \ref{subsec:rel:coir}.  

On the other hand, we do not seem to recover exactly corotational flow with $V^i = 0$ for $\eta = 1$ for the decomposition (\ref{rel:V_ans}), since it is not clear how, for $\eta = 1$, we would obtain a solution for $\Phi$ that exactly cancels the other terms in (\ref{rel:V_ans}).  As we have seen in Section \ref{subsec:Newt:coir}, we do obtain corotational flow for $\eta = 1$ in the Newtonian limit; for relativistic solutions, deviations from corotation can therefore be at most at the post-Newtonian order.   Our decomposition (\ref{rel:u_ans}) therefore represents a sequence that connects irrotational flow with {\em approximately} corotational flow.  In Appendix \ref{AppA} we present an alternative decomposition of the velocity field, that does reproduce exactly corotational flow for $\eta = 1$.  However, this decomposition also introduces several additional derivative terms that may prove problematic, as we discuss in Appendix \ref{AppA}.

\section{Summary and Discussion}
\label{sec:Sum}

In this paper we propose a formalism for the construction of binary stars with intermediate circulation.  Adopting the approach of \cite{Shi98} we generalize the equations for irrotational fluid flow and show that stars with intermediate circulation can be constructed by solving one additional elliptic equation.   We also propose a decomposition of the fluid flow that represents a sequence connecting irrotational stars with corotational ones in the Newtonian case, and approximately corotational ones in the relativistic case.   Exploring the physical properties of these sequences will require future numerical work.  It should also be possible to generalize the decomposition of the fluid flow to allow for rotation about other axes of rotation, or to modify the degree of differential rotation.   Our approach may also be used to construct neutron stars in quasiequilibrium orbit about black holes.

Relaxing the assumption of either corotational or irrotational flow in binary stars is, at the very least, interesting as a point of principle.  Such a formalism is necessary, however, for the construction of 
binaries containing rapidly rotating stars, for example millisecond pulsars, when the binary's orbital frequency is no longer much larger than the spin frequency of the star.  As suggested by MS, approximate inspiral sequences of such binaries can then be constructed by fixing the circulation along one particular fluid trajectory, for example along the stellar equator.   This can be achieved by iterating, at a given binary separation, over the parameter $\eta$ in the decomposition of the fluid flow until the desired value of the circulation has been realized.

\acknowledgments

This work was supported in part by NSF grant  PHY-0756514 to Bowdoin College,  NASA grants NNG04GK54G and NNX07AG96G to UIUC, and NSF grants PHY-0650377, PHY-0345151 and PHY-0205155 to UIUC.

\appendix

\section{An alternative decomposition of the fluid velocity}
\label{AppA}

As we have seen in Section \ref{subsec:rel:limits}, the decomposition (\ref{rel:u_ans}) does not necessarily lead to exactly corotational fluid flow in the limit $\eta = 1$.  An attractive alternative to this decomposition is
\begin{equation} \label{app:u_ans}
\hat u_i = D_i \Phi + \eta u^t h B_i,
\end{equation}
which leads to
\begin{equation}
V^i = \frac{1}{u^t h} D^i \Phi + (\eta - 1) B^i.
\end{equation}
Retracing the derivations of Section \ref{subsec:rel:arb_spin} we find that equation (\ref{rel:ut}) can now be written
\begin{equation}
u^t h = \frac{H + B^i D_i \Phi}{\alpha^2 - \eta B^i B_i},
\end{equation}
while the continuity equation (\ref{rel:cont}) becomes
\begin{eqnarray} \label{app:Phi_eq}
D^2 \Phi & + & (\eta - 1) \, D_i \left( \frac{H + B^j D_j \Phi}{\alpha^2 - \eta B^k B_k} B^i \right) 
\\
& = & \left((1 - \eta) \frac{H + B^j D_j \Phi}{\alpha^2 - \eta B^k B_k} B^i - D^i \Phi \right) D_i \ln \frac{\alpha\rho_0}{h}. \nonumber 
\end{eqnarray}
Taking a divergence of the Euler equation (\ref{rel:Euler}) results in
\begin{equation} \label{app:H_eq}
D^2 H + \eta D^i (V^j \kappa_{ij}) = 0,
\end{equation}
where we have now defined
\begin{equation}
\kappa_{ij} \equiv D_j (u^t h B_i) - D_i (u^t h B_j).
\end{equation}

We first note that, for $\eta = 0$, the above equations reduce to those for irrotational flow as before.  For $\eta = 1$, equation (\ref{app:Phi_eq}), together with its boundary condition, is now solved by $\Phi = C_\Phi$, where $C_\Phi$ is a constant.  That implies that, unlike our original decomposition (\ref{rel:u_ans}), the alternative decomposition (\ref{app:u_ans}) does reproduce corotational flow with $V^i = 0$ in the limit $\eta = 1$.    While this is a very attractive feature, this decomposition also has disadvantages.   Note that the term $u^t h$ contains first derivatives of $\Phi$, meaning that the term $D^i \kappa_{ij}$ in equation (\ref{app:H_eq}) will introduce third derivatives of $\Phi$.  Using the decomposition (\ref{rel:u_ans}), on the other hand, we avoid these third-derivative terms.  The appearance of $u^t h$ in equation (\ref{rel:H_eq}) still introduces second derivatives of $\Phi$.  All of these terms disappear in the Newtonian limit, however, are therefore small, and hopefully do not affect the well-behavedness of the system.


\newpage
   
 \onecolumngrid
   
\begin{center}
{\large \bf Erratum}
\end{center}

It was kindly brought to our attention by E. Gourgoulhon that the formalism that we propose to construct binary neutron stars with arbitrary circulation solves only a subset of the equilibrium equations in general, but not all of them.

Specifically, our formalism considers only the divergence of the Euler equation for stationary equilibrium, Eq.~(13) in the Newtonian case and Eq.~(37) in the relativistic case.  However, for the Euler equation to be satisfied, both its 
divergence {\it and} its curl has to vanish.  For the limiting cases of corotational or irrotational fluid flow the curl of these equations {\it does} 
vanish identically, but this is not necessarily the case for intermediate 
circulation.

While our formalism still leads to valid binary initial data for 
fluid flow with intermediate circulation, these data are not 
necessarily in exact equilibrium for the Newtonian case or quasiequilibrium in
the relativistic case.  Given that the curl vanishes for 
both limiting cases, it is possible that the 
curl is small for intermediate configurations as well, meaning 
that these data may represent good approximations to 
equilibrium configurations.  It may also 
be possible to construct equilibrium (or quasiequilibrium) data by adding another 
contribution to the decomposition of the fluid velocity, Eq.~(23) in the 
Newtonian case and Eq.~(52) in the relativistic case, and solving for this 
contribution by forcing the curl of the Euler equation to vanish.

\end{document}